\documentclass[pra]{revtex4}
 \usepackage{amssymb} \usepackage{graphicx}

\begin{document}
 \title{Supergravity backgrounds and symmetry superalgebras}

\author{\"Umit Ertem}
 \email{umitertemm@gmail.com}
\address{School of Mathematics, The University of Edinburgh, James Clerk Maxwell Building, Peter Guthrie Tait Road, Edinburgh, EH9 3FD, United Kingdom\\}
 
\begin{abstract}

We consider the bosonic sectors of supergravity theories in ten and eleven dimensions which correspond to the low energy limits of string theories and M-theory. The solutions of supergravity field equations are known as supergravity backgrounds and the number of preserved supersymmetries in those backgrounds are determined by Killing spinors. We provide some examples of supergravity backgrounds which preserve different fractions of supersymmetry. An important invariant for the characterization of supergravity backgrounds is their Killing superalgebras which are constructed out of Killing vectors and Killing spinors of the background. After constructing Killing superalgebras of some special supergravity backgrounds, we discuss about the possibilities of the extensions of these superalgebras to include the higher degree hidden symmetries of the background.

\end{abstract}

\maketitle

\section{Introduction}

The unification of fundamental forces of nature is one of the
biggest aims in modern theoretical physics. Most promising
approaches for that aim includes the ten dimensional supersymmetric
string theories and their eleven dimensional unification called
M-theory. There are five different string theories in ten dimensions
which are type I, type IIA and IIB and heterotic $E_8\times E_8$ and
$SO(32)$ theories. However, some dualities called T-duality,
S-duality and U-duality between strong coupling and weak coupling
limits of these theories can be defined and these dualities can give
rise to one unified M-theory in eleven dimensions \cite{Green Schwarz Witten, Townsend1, Duff,  Becker Becker Schwarz}. The main common
property of these ten and eleven dimensional theories is that their
low energy limits correspond to the supergravity theories in those
dimensions. This is the main reason for supergravity theories to
attract increasing attention in recent research literature \cite{Van Proeyen, Freedman Van Proeyen}.
Understanding supergravity theories is important for knowing the
dynamics of massless fields in string theories and finding the
backgrounds that strings can propagate.

Supergravity theories are extensions of General Relativity to obtain
an action that is invariant under supersymmetry transformations. For
the consistency and invariance of the theory one has to define
higher spin fermionic fields and extra bosonic fields with
appropriate supersymmetry transformations. There are different
consistent supergravity theories in different dimensions. The low
energy limits of string and M-theories are described by the bosonic
sectors of ten and eleven dimensional supergravity theories. Bosonic
sectors of supergravity theories correspond to taking fermionic
fields and their variations to be zero in the full theory. The
solutions of bosonic supergravity field equations are called
supergravity backgrounds. An important invariant that is used in the
classification of supergravity backgrounds is the number of
preserved supersymmetries in those backgrounds. The number of
supersymmetries are described by Killing spinors of the background
which correspond to the spinors that are solutions of the
differential equation results from the variation of the gravitino
field \cite{Townsend2}. This differential equation also defines the spinor covariant
derivative in terms of the extra bosonic fields of the theory.

The amount of supersymmetry in a supergravity background is the main tool for the classification of supersymmetric supergravity backgrounds. Constructing the Killing superalgebras of a background is an important step to reach this classification. A Killing superalgebra is a Lie superalgebra that consists of the isometries of the background which correspond to the Killing vectors and the number of preserved supersymmetries that constitutes the Killing spinors of the background. These constituents of the superalgebra are even and odd parts of it respectively. In the definition of Killing superalgebras, the spinorial Lie derivative and Dirac currents that are constructed from two Killing spinors and correspond to Killing vectors are used \cite{O Farrill Meessen Philip, O Farrill Jones Moutsopoulos}. To have a well defined Lie superalgebra structure, the Jacobi identities of odd and even parts of the superalgebra must be satisfied. The dimension of the Killing superalgebra is related to the homogeneity structure of the background \cite{O Farrill Hustler1, O Farrill Hustler2}. By constructing the Killing superalgebras of different supergravity backgrounds, one can obtain the geometric and supersymmetric properties of those backgrounds. On the other hand, classification of supergravity backgrounds can be considered in different contexts. Besides the construction of Killing superalgebras, there are also methods that use $G$-structures and spinorial geometry to attack on the classification problem \cite{Gauntlett Pakis, Gauntlett Gutowski Pakis, Gauntlett Martelli Pakis Waldram, Gran Papadopoulos Roest, Gran Gutowski Papadopoulos Roest}. However, we will focus on the method of constructing Killing superalgebras in the rest of the paper.

In this review paper, we consider the current status of the subject and discuss about the possibble extensions of the structures, which can give hints about how to achive some developments regarding the problem. The paper is organized as follows. In section 2, we summarize the bosonic sectors of different supergravity theories in ten and eleven dimensions which are the low energy limits of string theories and M-theory. In section 3, we provide some examples of solutions of those supergravity theories that are called supersymmetric supergravity backgrounds. Section 4 contains the construction and properties of Killing superalgebras in general supergravity backgrounds and some possible extensions of them to higher order forms. In section 5, we provide the summary and discussion before concluding the paper.

\section{Bosonic sectors of supergravity theories}

The low energy limits of string theories and M-theory are ten and
eleven dimensional supergravity theories. To obtain the solutions
that strings can propagate consistently, we consider the bosonic
sectors of supergravities. These correspond to the backgrounds of
the theories by taking the fermionic fields to be zero and
supersymmetry variations of fermionic fields give rise to Killing
spinor equations whose solutions give the number of supersymmetries
preserved by the background. In general, a supergravity background
consists of a Lorentzian spin manifold $M$, a metric $g$ and some
other bosonic fields defined on $M$ depending on the type and
dimension of the supergravity theory. The metric and bosonic fields
satisfy some field equations generalizing the Einstein-Maxwell
equations. A supergravity background is called supersymmetric if it
admits nonzero Killing spinors that are solutions of the Killing
spinor equation. In this section, we will introduce the bosonic sectors of supergravity theories in ten and eleven dimensions.

\subsection{Eleven-dimensional supergravity}

The maximum dimension that a supergravity theory can be consistently
constructed is eleven and eleven dimensional supergravity is the low
energy limit of M-theory \cite{Cremmer Julia Scherk}. Bosonic sector of eleven-dimensional
supergravity consists of a Lorentzian metric $g$ and a closed 4-form
$F$ which is the field strength of a 3-form $A$, namely $F=dA$. This
theory can be considered as a generalization of Einstein-Maxwell
theory to eleven dimensions with a generalized field strength $F$.
The action of the theory is written as follows
\begin{equation}
S=\frac{1}{12\kappa_{11}^2}\int\left(R_{ab}\wedge*e^{ab}-\frac{1}{2}F\wedge*F-\frac{1}{6}A\wedge F\wedge F\right)
\end{equation}
where $\kappa_{11}$ is the eleven dimensional gravitational coupling constant, $R_{ab}$ are curvature 2-forms, $e^a$ are coframe basis constructed from the metric $g$ with $e^{ab}=e^a\wedge e^b$ and $*$ is the Hodge star operation. Note that the second term in the action is in the form of the Maxwell action and the last term is the topological Chern-Simons term. Varying with respect to $e^a$ and $A$ gives rise to the following field equations;
\begin{eqnarray}
Ric(X,Y)*1&=&\frac{1}{2}i_XF\wedge*i_YF-\frac{1}{6}g(X,Y)F\wedge*F\\
d*F&=&\frac{1}{2}F\wedge F
\end{eqnarray}
where $X,Y$ are vector fields, $Ric$ is the Ricci tensor, $*1$ is the volume form and $i_X$ is the interior derivative or contraction with respect to $X$. On the other hand, the spinor covariant derivative is modified by the existence of the extra bosonic fields in supergravity theories. Variation of the gravitino field gives a spinor equation whose solutions are Killing spinors
\begin{equation}
{\cal{D}}_X\epsilon=0
\end{equation}
with $\epsilon$ is a spinor and the modified spinor covariant derivative is defined in terms of ordinary spinor covariant derivative $\nabla_X$ as
\begin{equation}
{\cal{D}}_X=\nabla_X+\frac{1}{6}i_XF-\frac{1}{12}\tilde{X}\wedge F
\end{equation}
and $\tilde{X}$ is the 1-form that corresponds to the metric dual of $X$. The solutions of the Killing spinor equation determine the number of preserved supersymmetries in a supergravity background.

\subsection{Type IIA and IIB supergravities}

By compactifying eleven-dimensional supergravity on $S^1$ with the
Kaluza-Klein reduction method, one obtains the ten-dimensional type
IIA supergravity which is the low energy limit of type IIA string
theory \cite{Duff Nilsson Pope}. Besides the fermionic fields gravitino and dilatino, type
IIA supergravity includes several bosonic fields that are the
graviton described by the metric $g$, a real scalar field called
dilaton $\phi$, a 2-form gauge potential $B_2$ with field strength
$H_3=dB_2$ and 1-form and 3-form Ramond-Ramond gauge potentials
$C_1$ and $C_3$ with field strengths $F_2=dC_1$ and $F_4=dC_3$. The
bosonic sector of IIA supergravity correspond to taking the
fermionic fields to be zero.

By defining a new field strength
\begin{equation}
\tilde{F}_4=F_4-C_1\wedge H_3
\end{equation}
with the property
\begin{equation}
d\tilde{F}_4=-F_2\wedge H_3
\end{equation}
the bosonic action of IIA supergravity is written as follows
\begin{eqnarray}
S&=&\frac{1}{2\kappa_{10}^2}\int e^{-2\phi}\left(R_{ab}\wedge*e^{ab}+4d\phi\wedge*d\phi\right)\nonumber\\
&&-\frac{1}{4\kappa_{10}^2}\int\left(e^{-2\phi}H_3\wedge*H_3+F_2\wedge*F_2+\tilde{F}_4\wedge*\tilde{F}_4\right)\nonumber\\
&&-\frac{1}{4\kappa_{10}^2}\int B_2\wedge F_4\wedge F_4.
\end{eqnarray}

The Killing spinors in type IIA supergravity satisfy two equations since there are two fermionic fields in the theory. The differential condition comes from the variation of the gravitino and the equation corresponds to the condition of being parallel with respect to the following modified spinor covariant derivative
\begin{equation}
{\cal{D}}_X=\nabla_X-\frac{1}{4}(i_XH_3)z-\frac{1}{8}e^{\phi}(i_XF_2)z+\frac{1}{8}e^{\phi}\tilde{X}\wedge F_4
\end{equation}
where $z$ is the ten dimensional volume form. The algebraic condition comes from the variation of the dilatino and reads as follows
\begin{equation}
\left(-\frac{1}{3}(d\phi)z+\frac{1}{6}H_3-\frac{1}{4}e^{\phi}F_2+\frac{1}{12}e^{\phi}F_4z\right)\epsilon=0.
\end{equation}

There is another ten dimensional supergravity theory which is called type IIB supergravity and can not be obtained from the eleven dimensional theory \cite{DallAgata Lechner Tonin}. Fermionic content of IIB theory is same as in IIA case, however the bosonic sector differs from it. Besides the metric $g$, dilaton $\phi$ and the 2-form $B_2$ with field strength $H_3=dB_2$, it also contains 0-form, 2-form and 4-form Ramond-Ramond gauge potentials $C_0$, $C_2$ and $C_4$ with field strengths $F_1=dC_0$, $F_3=dC_2$ and self-dual $F_5=dC_4=*F_5$. By defining the new combinations of field strengths
\begin{eqnarray}
\tilde{F_3}&=&F_3-C_0\wedge H_3\nonumber\\
\tilde{F_5}&=&F_5-\frac{1}{2}C_2\wedge H_3+\frac{1}{2}B_2\wedge F_3,
\end{eqnarray}
the action of the bosonic sector of the type IIB supergravity can be written as
\begin{eqnarray}
S&=&\frac{1}{2\kappa_{10}^2}\int e^{-2\phi}\left(R_{ab}\wedge*e^{ab}+4d\phi\wedge*d\phi\right)\nonumber\\
&&-\frac{1}{4\kappa_{10}^2}\int\left(e^{-2\phi}H_3\wedge*H_3+F_1\wedge*F_1+\tilde{F}_3\wedge*\tilde{F}_3+\tilde{F}_5\wedge*\tilde{F}_5\right)\nonumber\\
&&-\frac{1}{4\kappa_{10}^2}\int C_4\wedge H_3\wedge F_3.
\end{eqnarray}
In fact, it is only a pseudo-action since only after externally imposing the self-duality condition $F_5=*F_5$, the field equations of the theory can be obtained from it.

The differential Killing spinor equation in type IIB supergravity is written from the following modified spinor covariant derivative \cite{Jones Smith}
\begin{equation}
{\cal{D}}_X=\nabla_X+\frac{1}{4}i_XH_3\otimes\sigma_3-\frac{1}{16}e^{\phi}\tilde{X}\wedge\left(F_1\otimes i\sigma_2-\tilde{F_3}\otimes\sigma_1+\tilde{F_5}\otimes i\sigma_2\right)
\end{equation}
where $\sigma_i$ are Pauli matrices and the algebraic Killing spinor equation is
\begin{equation}
\left(\frac{1}{2}(d\phi-ie^{\phi}dC_0)+\frac{1}{4}(ie^{\phi}\tilde{F_3}-H_3)\right)\epsilon=0.
\end{equation}

\subsection{Type I and heterotic supergravities}

The low energy limit of type I superstring theory is a ten-dimensional supergravity theory coupled with a super Yang-Mills theory with gauge group $SO(32)$ in ten dimensions \cite{Chapline Manton}. This low energy limit is called as type I supergravity theory and it contains the metric $g$, the dilaton $\phi$, a Ramond-Ramond 2-form $C_2$ and the non-abelian Yang-Mills $SO(32)$ gauge connection $A_1$  as bosonic fields. The field strengths are defined as
\begin{eqnarray}
F_2&=&dA_1+A_1\wedge A_1\nonumber\\
\tilde{F_3}&=&dC_2+\frac{1}{4}\omega_3\nonumber
\end{eqnarray}
where $\omega_3=A_1\wedge dA_1+\frac{2}{3}A_1\wedge A_1\wedge A_1$ is the Chern-Simons form of the gauge connection. The bosonic action of type I supergravity is defined in terms of these field strengths as follows
\begin{equation}
S=\int\left[e^{-2\phi}\left(R_{ab}\wedge*e^{ab}+4d\phi\wedge*d\phi\right)-\frac{1}{2}\tilde{F_3}\wedge*\tilde{F_3}-e^{-\phi}F_2\wedge*F_2\right].
\end{equation}
The variation of the gravitino gives rise to the following spinor covariant derivative which is used in the definition of the Killing spinors;
\begin{equation}
{\cal{D}}_X=\nabla_X-\frac{1}{8}e^{\phi}\tilde{F_3}\tilde{X}.
\end{equation}
Besides the gravitino, there are two more fermionic fields which are dilatino and gaugino and hence we have the following two algebraic Killing spinor equations
\begin{eqnarray}
\left(d\phi+\frac{1}{2}e^{\phi}\tilde{F_3}\right)\epsilon&=&0\nonumber\\
F_2\epsilon&=&0.
\end{eqnarray}

The heterotic supergravity differs from type I supergravity in its fermionic sector. Since we consider only the bosonic sectors of supergravities, we can write the bosonic action of the heteroric supergravity in terms of the metric $g$, the dilaton $\phi$, a 2-form $B_2$ with field strength $H_3=dB_2$ and 1-form non-abelian $SO(32)$ or $E_8\times E_8$ gauge potential $A_1$ with the field strength $F_2$ as in the type I case;
\begin{equation}
S=\int e^{-2\phi}\left(R_{ab}\wedge*e^{ab}+4d\phi\wedge*d\phi-\frac{1}{2}H_3\wedge*H_3-\frac{1}{2}F_2\wedge*F_2\right).
\end{equation}
The differential Killing spinor equation is determined by the following spin connection
\begin{equation}
{\cal{D}}_X=\nabla_X-\frac{1}{4}i_XH_3.
\end{equation}
 Although the fermionic sector of heterotic and type I supergravities are not same, the algebraic Killing spinor equations of heterotic supergravity has a similar form as in the type I case;
\begin{eqnarray}
\left(d\phi+\frac{1}{2}H_3\right)\epsilon&=&0\nonumber\\
F_2\epsilon&=&0.
\end{eqnarray}

\section{Supergravity backgrounds}

There are various special solutions for eleven-dimensional and ten-dimensional supergravity theories. One of the methods that can give hints about obtaining all solutions is to find a way of classification for them. However, the complete classification of all supergravity backgrounds has not been achieved yet. On the other hand, for some special cases such as for backgrounds that have maximal supersymmetry and for symmetric space backgrounds, the classification problem can be accomplished. In this section, we consider some special solutions of the supergravity theories and summarize the classification results for symmetric and maximally supersymmetric backgrounds.

In eleven dimensions, spinor space is a 32-dimensional real space and hence the space of the solutions of Killing spinor equation can be at most 32-dimensional. An important invariant for a supersymmetric supergravity background is the fraction of preserved supersymmetries in that background which is denoted as $\nu=\frac{1}{32}n$ and $n$ is the dimension of the Killing spinor space. There are various numbers of known supergravity backgrounds which have different fractions of preserved supersymmetries. Indeed, the known solutions correspond to $n=0,1,2,3,4,5,6,...,8,...,12,...,16,...,18,...,20,...,22,...,24,...,26,...,32$ \cite{O Farrill Meessen Philip}. The numbers that does not appear in the list correspond to the cases of the fraction of preserved supersymmetries whose exact forms are not known. Moreover, for $n=30$ and $31$, it is known that there is no supersymmetric supergravity backgrounds \cite{Gran Gutowski Papadopoulos, Bandos Azcarraga Varela, Gran Gutowski Papadopoulos Roest2, O Farrill Gadhia}. It is also proved that for the cases of $\nu>\frac{1}{2}$, those supergravity backgrounds are locally homogeneous \cite{O Farrill Hustler1, O Farrill Hustler2}.  For $\nu=1$, the solutions are called maximally supersymmetric or BPS solutions and examples for this case includes eleven-dimensional flat Minkowski spacetime and Freund-Rubin backgrounds such as $AdS_7\times S^4$ and $AdS_4\times S^7$. There are also half-BPS, namely $\nu=\frac{1}{2}$, solutions of the eleven-dimensional supergravity. Examples for this case are M-wave, Kaluza-Klein monopole and M2- and M5-brane solutions \cite{Acharya O Farrill Hull Spence}.

\subsection{Half-BPS solutions}

A gravitational $pp$-wave is defined as a spacetime with a parallel null vector. A supersymmetric gravitational $pp$-wave which is called the M-wave is a solution of eleven dimensional supergravity and its metric is given as follows \cite{Hull}
\begin{equation}
ds^2=2dx^+dx^-+a(dx^+)^2+(dx^9)^2+g_{ij}dx^idx^j,
\end{equation}
where $i,j=1$ to $8$, $x^+,x^-$ are light cone coordinates, $a$ is an arbitrary function with $\partial_-a=0$ and $g_{ij}$ is a family of metrics dependent on $x^+$. The holonomy group of the manifold on which $g_{ij}$ is defined is contained in $Spin(7)$ and the following property is satisfied
\begin{equation}
\partial_+\Omega=\lambda\Omega+\Psi
\end{equation}
with $\Omega$ is the self-dual $Spin(7)$-invariant Cayley 4-form, $\lambda$ a smooth function of $(x^+,x^-)$ and $\Psi$ is a anti self-dual 4-form. The closed 4-form $F$ of eleven-dimensional supergravity vanishes in this solution. The metric (21) is a supersymmetric solution of eleven-dimensional supergravity if and only if it is Ricci-flat. By dropping the $x^9$ coordinate, one can also obtain a solution of ten-dimensional supergravity which is written in the following form
\begin{equation}
ds^2=2dx^+dx^-+a(dx^+)^2+g_{ij}dx^idx^j.
\end{equation}

There is also a brane solution of eleven-dimensional supergravity with the following metric which describes a number of parallel M2-branes \cite{Duff Stelle};
\begin{equation}
g=H^{-2/3}g_{2+1}+H^{1/3}g_8
\end{equation}
where $g_{2+1}$ is the metric on the three-dimensional Minkowskian woldvolume $E^{2,1}$ of the branes and $g_8$ is the metric on the eight-dimensional Euclidean space $E^8$ transverse to the branes. $H$ is a harmonic function on $E^8$ and can be chosen as
\begin{equation}
H(r)=1+\frac{a^6}{r^6}
\end{equation}
with $r$ is the radial coordinate and $a^6=2^5\pi^2N{l_p}^6$. Here
$N$ is the number of coincident membranes at $r=0$ and $l_p$ is the
eleven-dimensional Planck length. The harmonic function has the
property $\lim_{r\rightarrow\infty}H(r)=1$. The closed 4-form field
in this background corresponds to
\begin{equation}
F=\pm z_{2+1}\wedge dH^{-1}
\end{equation}
where $z_{2+1}$ is the volume form of the brane worldvolume. The
M2-brane solution preserves half of the supersymmetries namely
$\nu=\frac{1}{2}$. However, it interpolates between two maximally
supersymmetric solutions; near the brane horizon $r\ll a$, it
corresponds to the maximally supersymmetric Freund-Rubin background
$AdS_4\times S^7$ and for infinitely far away from the brane
$r\rightarrow\infty$, it corresponds to the flat Minkowski space
$E^{10,1}$.

There is a similar brane solution describing a number of parallel M5-branes with metric and 4-form are given as follows \cite{Guven}
\begin{eqnarray}
g&=&H^{-1/3}g_{5+1}+H^{2/3}g_5\\
F&=&\pm3*_5H
\end{eqnarray}
where $g_{5+1}$ is the metric on the six-dimensional Minkowskian
worldvolume $E^{5,1}$ of the branes, $g_5$ and $*_5$ is the metric
and Hodge dual on the five-dimensional Euclidean space $E^5$
transverse to the branes. The harmonic function $H$ is defined as
\begin{equation}
H(r)=1+\frac{a^3}{r^3}
\end{equation}
where $a^3=\pi N{l_p}^3$, $N$ is the number of coincident fivebranes
at $r=0$ and $\lim_{r\rightarrow\infty}H(r)=1$. M5-brane preserves
half of the supersymmetries and it interpolates between the
maximally supersymmetric Freund-Rubin background $AdS_7\times S^4$
and flat Minkowski space $E^{10,1}$.

The ten-dimensional type IIB supergravity also has a brane solution which is called D3-brane with the metric given by \cite{Horowitz Strominger}
\begin{equation}
g=H^{-1/2}g_{3+1}+H^{1/2}g_6
\end{equation}
where $g_{3+1}$ is the metric on the worldvolume $E^{3,1}$ of the brane and $g_6$ is the Euclidean metric on the transverse space $E^6$ to the brane. The self-dual 5-form $F_5$ is defined on $S^5\subset E^6$ and has quantized flux and the dilaton $\phi$ is constant. The harmonic function $H$ is defined as
\begin{equation}
H(r)=1+\frac{a^4}{r^4}
\end{equation}
with $a^4=4\pi gN{l_s}^4$ and $g$ is the string coupling constant and $l_s$ is the string length. $N$ is the number of parallel D3-branes at $r=0$. This solution interpolates between flat Minkowski space at infinity and $AdS_5\times S^5$ at the near horizon limit.

\subsection{Maximally supersymmetric backgrounds}

The solutions of supergravity theories which have maximum number of Killing vector fields and maximum number of Killing spinors are called maximally supersymmetric supergravity backgrounds. In this case the dimension of the space of Killing spinors is 32 and we have $\nu=1$.

In eleven dimensions, there are four classes of solutions that have maximal supersymmetry. First one is the flat Minkowski spacetime $M^{1,10}$ with $F=0$. Two of the other non-trivial classes include the Freund-Rubin backgrounds of which case we have a four-dimensional and seven-dimensional split of spacetime $M=M^4\times M^7$ and the total metric is written as the sum of the metrics of the split spacetimes;
\begin{equation}
ds^2={ds_4}^2+{ds_7}^2.
\end{equation}
One of the split spacetimes has positive and the other one has negative constant curvatures. First case of this type of solutions corresponds to the following background;
\begin{eqnarray}
AdS_7(-7R)\times S^4(8R)\nonumber\\
F=\sqrt{6R}z_{S^4}
\end{eqnarray}
where the numbers in paranthesis correspond to the constant scalar curvatures of the relevant backgrounds with $R>0$ and $z_{S^4}$ is the volume form of $S^4$. The second case is the following spacetime;
\begin{eqnarray}
AdS_4(8R)\times S^7(-7R)\nonumber\\
F=\sqrt{-6R}z_{AdS^4}
\end{eqnarray}
with $R<0$ and $z_{AdS^4}$ is the volume form of $AdS_4$. The fourth class of maximally supersymmetric solutions is a one parameter family of symmetric plane waves with the metric and the 4-form
\begin{eqnarray}
g&=&2dx^+dx^--\frac{1}{36}\mu^2\left(4\sum_{i=1}^3(x^i)^2+\sum_{i=4}^9(x^i)^2\right)(dx^-)^2+\sum_{i=1}^9(dx^i)^2\nonumber\\
F&=&\mu dx^-\wedge dx^1\wedge dx^2\wedge dx^3
\end{eqnarray}
where $\mu$ is a real number.

In ten dimensions, the only maximally supersymmetric solution of type I, heterotic and IIA supergravities is the flat Minkowski spacetime. However, in type IIB case we have two non-trivial classes. The first one is the Freund-Rubin background
\begin{eqnarray}
AdS_5(-R)\times S^5(R)\nonumber\\
F_5=\sqrt{\frac{4R}{5}}\left(z_{AdS_5}-z_{S^5}\right)
\end{eqnarray}
where $F_5$ is the self-dual 5-form and $R>0$. The second one is a family of symmetric plane waves with the following metric and self-dual 5-form $F_5$
\begin{eqnarray}
g&=&2dx^+dx^--\frac{1}{4}\mu^2\sum_{i=1}^8(x^i)^2(dx^-)^2+\sum_{i=1}^8(dx^i)^2\nonumber\\
F_5&=&\frac{1}{2}\mu dx^-\wedge\left(dx^1\wedge dx^2\wedge dx^3\wedge dx^4+dx^5\wedge dx^6\wedge dx^7\wedge dx^8\right).
\end{eqnarray}

The classification of all supergravity backgrounds is one of the main goals of the study of supergravity theories. The homogeneity theorem for supergravity backgrounds says that a supergravity background which preserve more then half of the supersymmetries must be homogeneous \cite{O Farrill Hustler1, O Farrill Hustler2}. A homogeneous spacetime can be characterized as a spacetime of which the tangent space at any point is spanned by Killing vectors that are constructed by squaring the Killing spinors of the background. If a homogeneous background also corresponds to a symmetric space, then we call it a symmetric supergravity background. The classification of symmetric supergravity backgrounds in eleven-dimensional and ten-dimensional type IIB cases are achieved in \cite{O Farrill Papadopoulos}. However, the full classification of all supergravity backgrounds is still an open problem.

\section{Killing superalgebras}

An important invariant that characterizes the supersymmetric
supergravity backgrounds is the Killing superalgebra of that
background \cite{O Farrill Meessen Philip, O Farrill}. A Killing superalgebra $\mathfrak{g}$ has a Lie
superalgebra structure which consists of a $\mathbb{Z}_2$-graded
algebra that is a direct sum of two components
$\mathfrak{g}=\mathfrak{g}_0\oplus\mathfrak{g}_1$. The first
component $\mathfrak{g}_0$ of the Lie superalgebra is called the even part and has a Lie algebra structure, the
second one $\mathfrak{g}_1$ is called the odd part and corresponds to a module of $\mathfrak{g}_0$. A Lie bracket $[ , ]$ on a Lie superalgebra is defined as a bilinear multiplication
\begin{equation}
[ , ]:\mathfrak{g}_i\times\mathfrak{g}_j\longrightarrow\mathfrak{g}_{i+j}
\end{equation}
where $i, j=0, 1$ and satisfies the following skew-supersymmetry and super-Jacobi identities for $a, b, c$ are elements of $\mathfrak{g}$ and $|a|$ denotes the degree of $a$ which corresponds to 0 or 1 depending on that $a$ is in $\mathfrak{g}_0$ or $\mathfrak{g}_1$ respectively
\begin{eqnarray}
[a,b]&=&-(-1)^{|a||b|}[b,a]\nonumber\\
\left[a,[b,c]\right]&=&[[a, b], c]+(-1)^{|a||b|}[b, [a, c]].
\end{eqnarray}
 For a Killing superalgebra, the
even and odd parts of it are defined as bosonic and fermionic parts
of the superalgebra. The bosonic part of the Killing superalgebras
of supergravity backgrounds corresponds to the Lie algebra of Killing
vector fields which are generated by Killing spinors. A Killing vector field $K$ is an isometry of the background which means that the Lie derivative of the metric $g$ with respect to $K$ is zero
\begin{equation}
{\cal{L}}_Kg=0.
\end{equation}
Killing vector fields in the Killing superalgebra also preserve the other bosonic fields in the corresponding supergravity background. The fermionic part of the Killing superalgebra consists of the Killing spinors
of the background. The supersymmetric properties of a supergravity
background can be found from the Killing superalgebra which consists
of the supersymmetry generators and isometries of the background.

Three operations corresponding to the Lie brackets in the
superalgebra can be defined for the even (bosonic) and odd
(fermionic) parts of a Killing superalgebra. Since the even part
corresponds to the algebra of Killing vector fields, the Lie bracket
defined on it is the ordinary Lie bracket for vector fields
\begin{equation}
[\,,\,]:\mathfrak{g}_0\times\mathfrak{g}_0\longrightarrow\mathfrak{g}_0.
\end{equation}
The action of the even part to the odd part is defined as the Lie derivative of spinor fields with respect to a Killing vector;
\begin{equation}
{\cal{L}}:\mathfrak{g}_0\times\mathfrak{g}_1\longrightarrow\mathfrak{g}_1.
\end{equation}
The Lie derivative of a Killing spinor will be again a Killing spinor. Spinor Lie derivative is defined only with respect to Killing vectors and it is induced from the Lie derivative on differential forms as sections of the Clifford bundle of the manifold \cite{Lichnerowicz, Kosmann}. It is written in terms of the spinor covariant derivative and a 2-form constructed from a Killing 1-form as follows
\begin{equation}
{\cal{L}}_K\psi=\nabla_K\psi+\frac{1}{4}(d\tilde{K})\psi
\end{equation}
where $\tilde{K}$ is the Killing 1-form that is the metric dual of the Killing vector $K$. The third operation which takes two odd elements to obtain an even element is described by the squaring map of spinors
\begin{equation}
\mathfrak{g}_1\times\mathfrak{g}_1\longrightarrow\mathfrak{g}_0.
\end{equation}
This map is defined in terms of the spin invariant inner product on spinors \cite{Benn Tucker}. The Clifford product of a spinor $\psi$ with a dual spinor $\bar{\phi}$ can be written as a sum of differential forms by using the spinor inner product $( , )$. This decomposition is known as the Fierz identity;
\begin{equation}
\psi\bar{\phi}=(\psi,\phi)+(\psi,e_a\phi)e^a+(\psi,e_{a_2a_1}\phi)e^{a_1a_2}+...+(-1)^{\lfloor\frac{n}{2}\rfloor}(\psi,z\phi)z
\end{equation}
where $\lfloor\rfloor$ is the floor function and $z$ is the volume form. The squaring map corresponds to the projecting the Fierz identity onto the 1-form component. If $\psi$ and $\phi$ are Killing spinors, then the dual of the resulting 1-form is a Killing vector and for $\psi=\phi$, it is called the Dirac current $V_{\psi}$ of $\psi$.

In this way, all the needed brackets are defined for the Killing superalgebra. However, to obtain a Lie superalgebra structure, the Jacobi identities of the algebra must be satisfied. There are four Jacobi identities correspond to the $[\mathfrak{g}_0,\mathfrak{g}_0,\mathfrak{g}_0]$, $[\mathfrak{g}_0,\mathfrak{g}_0,\mathfrak{g}_1]$, $[\mathfrak{g}_0,\mathfrak{g}_1,\mathfrak{g}_1]$ and $[\mathfrak{g}_1,\mathfrak{g}_1,\mathfrak{g}_1]$ components. The first one is the ordinary Jacobi identity for the Lie algebra of Killing vector fields. The second and third ones are equivalent to the following properties of the spinor Lie derivative defined on spinors
\begin{equation}
[{\cal{L}}_{K_1},{\cal{L}}_{K_2}]\psi={\cal{L}}_{[K_1,K_2]}\psi
\end{equation}
\begin{equation}
{\cal{L}}_K(\psi\bar{\phi})=({\cal{L}}_K\psi)\bar{\phi}+\psi\overline{{\cal{L}}_K\phi}.
\end{equation}
The fourth Jacobi identity is the vanishing of the Lie derivative of a spinor with respect to the Dirac current of itself;
\begin{equation}
{\cal{L}}_{V_{\psi}}\psi=0.
\end{equation}
This is not a trivial result and it has been proved for different cases in \cite{O Farrill Meessen Philip, O Farrill Jones Moutsopoulos}.

To construct the Killing superalgebra of a supergravity background, one needs to know the isometry algebra of the background which consists of the Killing vectors on that background and the algebra consisting of the space of Killing spinors. However, by using the cone construction \cite{Bar}, which states that there is a one to one correspondence between the Killing spinors of a background and the parallel spinors of a background corresponding to the metric cone over the first one, one can also construct the Killing superalgebra without knowing the exact form of Killing spinors. For some special examples such as Freund-Rubin backgrounds in eleven dimensions, this kind of a construction can be achieved. For the $AdS_4\times S^7$ spacetime, the isometry algebra consists of the direct product of isometry algebras of the component spaces. Isometry algebra of $AdS_4$ is $\mathfrak{so}(3,2)$ and of $S^7$ is $\mathfrak{so}(8)$, so the isometry algebra of $AdS_4\times S^7$ is $\mathfrak{so}(3,2)\times\mathfrak{so}(8)$ and by using the cone construction procedures the Killing spinor space can also be obtained \cite{O Farrill}. As a result, the Killing superalgebra of this background is the following orthosymplectic Lie superalgebra;
\begin{equation}
\mathfrak{osp}(8|4).
\end{equation}
As another example, the isometry algebra of $AdS_7\times S^4$ is $\mathfrak{so}(6,2)\times\mathfrak{so}(5)$ and the Killing superalgebra is
\begin{equation}
\mathfrak{osp}(6,2|4).
\end{equation}
In eleven dimensional Minkowski background, the isometries generated by Killing spinors are translational Killing vector fields and the Killing superalgebra correspond to the supertranslation ideal of the Poincare superalgebra. However, one can also consider all isometries of the background that also contain rotational Killing vector fields and extend the Killing superalgebra to a symmetry superalgebra corresponding to the Poincare superalgebra. The Killing superalgebras of the fourth class of eleven dimensional maximally supersymmetric backgrounds denoted in (35) can be obtained by taking contractions of the Killing superalgebras of Freund-Rubin backgrounds.

Killing superalgebras play an important role in the classification of supergravity backgrounds. Maximally supersymmetric backgrounds which have $\nu=1$ and the minimally superymmetric ones can be classified completely. However, the classification of less than maximal and more than minimal supersymmetric bakgrounds is related to the Killing superalgebras of them. In eleven dimensions, the backgrounds preserving $\nu>\frac{1}{2}$ fraction of supersymmetry are locally homogeneous and this result is achieved by first constructing the Killing superalgebras and then obtaining the dimension of the traslational component of the squaring map from Killing spinors to Killing vectors. In ten dimensions, local homogeneity is guaranteed for $\frac{1}{2}\leq\nu\leq\frac{3}{4}$ and this can be proved from the construction of the Killing superalgebras \cite{O Farrill Jones Moutsopoulos}.

\subsection{Extension to higher order superalgebras}

Isometries of a background which are generated by Killing spinors constitute the even part of the Killing superalgebras. If one considers all isometries of the background (not necessarily generated by Killing spinors), then one can define more general symmetry superalgebras of the background. Besides this fact, the squaring map of spinors can be extended to define higher order spinor bilinears. Although there are some attempts to include these higher order objects in the symmetry superalgebras, the construction of the so-called maximal superalgebras is still an open problem \cite{O Farrill Jones Moutsopoulos Simon, Medeiros Hollands}.  It is known that the Killing vector fields have higher order antisymmetric generalizations to hidden symmetries and these hidden symmetries are called Killing-Yano (KY) forms. A KY $p$-form $\omega$ is defined as the solution of the following equation
\begin{equation}
\nabla_X\omega=\frac{1}{p+1}i_Xd\omega.
\end{equation}
Moreover, one can also define the generalizations of Dirac currents to higher-degree forms. From Fierz identity (45), one can define projection operators $\wp_p$ that projects onto the $p$-form component of the Clifford product of a spinor with its dual. The generalized currents are called $p$-form Dirac currents and defined as follows
\begin{equation}
\wp_p(\psi\bar{\psi})=(\psi,e_{a_p...a_2a_1}\psi)e^{a_1a_2...a_p}.
\end{equation}
It is shown in \cite{Acik Ertem} that these $p$-form Dirac currents of twistor spinors correspond to conformal Killing-Yano forms and for the Killing spinors case they correspond to the KY forms. It is also known that KY forms satisfy a graded Lie algebra structure in constant curvature space-times \cite{Kastor Ray Traschen}. The following Schouten-Nijenhuis bracket is defined for a $p$-form $\alpha$ and a $q$-form $\beta$
\begin{equation}
\left[\alpha, \beta\right]_{SN}=i_{X_a}\alpha\wedge\nabla_{X^a}\beta+(-1)^{pq}i_{X_a}\beta\wedge\nabla_{X^a}\alpha
\end{equation}
and corresponds to a Lie bracket for KY forms in constant curvature space-times. This means that the Killing superalgebras can be extended to include KY forms in some constant curvature supergravity backgrounds. On the other hand, a generalized Lie derivative on spinor fields with respect to KY forms have to be defined and the Jacobi identites also have to be satisfied. The spinorial Lie derivatives with respect to Killing vector fields also correspond to the symmetry operators of the Dirac equation in curved backgrounds. Symmetry operators of the Dirac equation that contain higher degree forms are also constructed in curved backgrounds by using KY forms \cite{Benn Charlton, Benn Kress, Acik Ertem Onder Vercin, Cariglia Krtous Kubiznak}. Hence, these operators are natural candidates for the generalized Lie derivatives of extended superalgebras. The construction of these extended Killing superalgebras can give rise to new hints about the classification problem of supergravity backgrounds. As a side remark, KY forms are also used in relation to $G$-structures in the supergravity context \cite{Papadopoulos}.

\section{Discussion}

Ten and eleven dimensional supergravity backgrounds correspond to the spacetimes that strings can propagate in a well defined manner. So, the complete classification of supergravity backgrounds can give hints about the possible string backgrounds and the unification of string theories. There are many known solutions of supergravity theories which correspond to the backgrounds that have different fraction of preserved supersymmetries. Finding the common geometrical properties of these backgrounds and the complete classification of them according to the preserved supersymmetries is one of the main problems in supergravity and string theory.  In some special cases, the classification problem is generally understood, however, in most cases it is not completely achieved yet. To achieve this aim, finding the invariants of these backgrounds is an important step. One of the main invariants of supersymetric supergravity backgrounds is their Killing superalgebras and they are constructed out of the isometries and Killing spinors of the background.

To construct a Killing superalgebra in a special supergravity background, one needs to know the Lie algebras of Killing vectors and Killing spinors. However, without knowing the Killing spinors of the background, one can also find the odd part of the superalgebra by using the cone construction and the knowledge about the parallel spinors. In this construction, the Lie derivatives of spinor fields and the Dirac currents of spinors are used and satisfying the Jacobi identities of the superalgebra completes the construction procedure. By this way, the Killing superalgebras of some supersymmetric backgrounds are obtained in the literature. Structure of these Killing superalgebras in different backgrounds and relations between them give a way to classify these supersymmetric backgrounds. For example, by using the properties of Killing superalgebras, the local homogeneity of a supergravity background that preserves more than half of supersymetries is proved \cite{O Farrill Hustler1, O Farrill Hustler2}.

Obtaining more hints about the classification problem can be possible by extending the Killing superalgebras to higher order geometric objects. Killing vector fields have natural antisymmetric generalizations to higher-degree forms which correspond to KY forms. Dirac currents also have generalizations to higher-degree components that are called $p$-form Dirac currents. The correspondence between the $p$-form Dirac currents and KY forms is proved in \cite{Acik Ertem}. So, extending the Killing superalgebras to include KY forms and $p$-form Dirac currents with a new definition of the generalized spinor Lie derivative can be possible. The properties of these extended superalgebras may give new insights to the classification of supergravity backgrounds.

\begin{acknowledgments}
The author thanks \"{O}zg\"{u}r A\c{c}{\i}k, Jose M. Figueroa-O`Farrill and Andrea Santi for useful discussions on the subject. This work is supported by the Scientific and Technological
Research Council of Turkey (T\"{U}B\.{I}TAK) grant B\.{I}DEB 2219.
\end{acknowledgments}

%\references%


\begin{references}

\bibitem{Green Schwarz Witten} Green, M. M.; Schwarz, J. H.; Witten, E. \emph{Superstring Theory}; vols. 1 and 2, Cambridge U. P., 1987.

\bibitem{Townsend1} Townsend, P. K. \emph{Four lectures on M-theory}, arxiv:hep-th/9612121.

\bibitem{Duff} Duff, M. J. \emph{ Int. J. Mod. Phys. A} \textbf{1996}, \emph{11}, 5623-5641.

\bibitem{Becker Becker Schwarz} Becker, K.; Becker, M.; Schwarz, J. H. \emph{String Theory and M-theory: A Modern Introduction}, Cambridge U. P., 2007.

\bibitem{Van Proeyen} Van Proeyen, A. \emph{Structure of supergravity theories}, arXiv:hep-th/0301005.

\bibitem{Freedman Van Proeyen} Freedman, D. Z.; Van Proeyen, A. \emph{Supergravity}, Cambridge U. P., 2012.

\bibitem{Townsend2} Townsend, P. K. In \emph{Novelties in String Theory: Proceedings of the Johns Hopkins Workshop on Current Problems in Particle Theory 22}, Goteborg, Sweden, 20-22 August 1998; Brink, L.; Marnelius, R., Eds; World Scientific Publ., Singapore, 1999, p.177.

\bibitem{O Farrill Meessen Philip} Figueroa-O'Farrill, J.; Meessen, P.; Philip, S. \emph{Classical Quant. Grav.} \textbf{2005}, \emph{22}, 207-226.

\bibitem{O Farrill Jones Moutsopoulos} Figueroa-O'Farrill, J.; Hackett-Jones, E.; Moutsopoulos, G. \emph{Classical Quant. Grav.} \textbf{2007}, \emph{24}, 3291-3308.

\bibitem{O Farrill Hustler1} Figueroa-O'Farrill, J.; Hustler, N. \emph{J. High Energy Phys.} \textbf{2012}, \emph{1210}, 014.

\bibitem{O Farrill Hustler2} Figueroa-O'Farrill, J.; Hustler, N. \emph{J. High Energy Phys.} \textbf{2014}, \emph{1404}, 131.

\bibitem{Gauntlett Pakis} Gauntlett, J. P.; Pakis, S. \emph{J. High Energy Phys.} \textbf{2003}, \emph{0304}, 039.

\bibitem{Gauntlett Gutowski Pakis} Gauntlett, J. P.; Gutowski, J. B.; Pakis, S. \emph{J. High Energy Phys.} \textbf{2003}, \emph{0312}, 049.

\bibitem{Gauntlett Martelli Pakis Waldram} Gauntlett, J. P.; Martelli, D.; Pakis, S.; Waldram, D. \emph{Commun. Math. Phys.} \textbf{2004}, \emph{247}, 421-445.

\bibitem{Gran Papadopoulos Roest} Gran, U.; Papadopoulos, G., Roest, D. \emph{Classical Quant. Grav.} \textbf{2005}, \emph{22}, 2701-2744.

\bibitem{Gran Gutowski Papadopoulos Roest} Gran, U.; Gutowski, J.; Papadopoulos, G., Roest, D. \emph{Classical Quant. Grav.} \textbf{2006}, \emph{23}, 1617-1678.

\bibitem{Cremmer Julia Scherk} Cremmer, E.; Julia, B.; Scherk, J. \emph{Phys. Lett. B} \textbf{1978}, \emph{76}, 409-412.

\bibitem{Duff Nilsson Pope} Duff, M. J.; Nilsson, B. E. W.; Pope, C. N. \emph{Phys. Rep.}
\textbf{1986}, \emph{130}, 1-142.

\bibitem{DallAgata Lechner Tonin} Dall`Agata, G.; Lechner, K.; Tonin, M. \emph{J. High Energy Phys.} \textbf{1998}, \emph{9807}, 017.

\bibitem{Jones Smith} Hackett-Jones, E. J.; Smith, D. J.; \emph{J. High Energy Phys.} \textbf{2004}, \emph{0411}, 029.

\bibitem{Chapline Manton} Chapline, G.; Manton, N. \emph{Phys. Lett. B} \textbf{1983}, \emph{120}, 105-109.

\bibitem{Gran Gutowski Papadopoulos} Gran, U.; Gutowski, J.; Papadopoulos, G. \emph{J. High Energy Phys.} \textbf{2007}, \emph{0702}, 044.

\bibitem{Bandos Azcarraga Varela} Bandos, I. A.; de Azcarraga, J. A.; Varela, O. \emph{J. High Energy Phys.} \textbf{2006}, \emph{0609}, 009.

\bibitem{Gran Gutowski Papadopoulos Roest2} Gran, U.; Gutowski, J.; Papadopoulos, G.; Roest, D. \emph{J. High Energy Phys.} \textbf{2007}, \emph{0702}, 043.

\bibitem{O Farrill Gadhia} Figueroa-O'Farrill, J.; Gadhia, S. \emph{J. High Energy Phys.} \textbf{2007}, \emph{0706}, 043.

\bibitem{Acharya O Farrill Hull Spence} Acharya, B. S.; Figueroa-O'Farrill, J. M.; Hull, C. M.; Spence, B. \emph{Adv. Theor. Math. Phys.} \textbf{1999}, \emph{2}, 1249-1286.

\bibitem{Hull} Hull, C. M.; \emph{Phys. Lett. B} \textbf{1984}, \emph{139}, 39-41.

\bibitem{Duff Stelle} Duff, M. J.; Stelle, K. S. \emph{Phys. Rep.} \textbf{1991}, \emph{253}, 113-118.

\bibitem{Guven} G\"{u}ven, R. \emph{Phys. Lett. B} \textbf{1992}, \emph{276}, 49-55.

\bibitem{Horowitz Strominger} Horowitz, G. T.; Strominger, A. \emph{Nucl. Phys. B} \textbf{1991}, \emph{360}, 197-209.

\bibitem{O Farrill Papadopoulos} Figueroa-O'Farrill, J. M.; Papadopoulos, G. \emph{J. High Energy Phys.} \textbf{2003}, \emph{0303}, 048.

\bibitem{O Farrill} Figueroa-O'Farrill, J. \emph{Classical Quant. Grav.} \textbf{1999}, \emph{16}, 2043-2056.

\bibitem{Lichnerowicz} Lichnerowicz, A. \emph{C. R. Acad. Sci. Paris} \textbf{1963}, \emph{257}, 7-9.

\bibitem{Kosmann} Kosmann, Y. \emph{Annal. Math. Pura ed Appl.} \textbf{1972}, \emph{91}, 317-395.

\bibitem{Benn Tucker} Benn, I. M.; Tucker, R. W. \emph{An Introduction to Spinors and Geometry with Applications in
Physics}, IOP Publishing Ltd, Bristol, 1987.

\bibitem{Bar} B\"{a}r, C. \emph{Commun. Math. Phys.} \textbf{1993}, \emph{154}, 509-521.

\bibitem{O Farrill Jones Moutsopoulos Simon} Figueroa-O'Farrill, J.; Hackett-Jones, E.; Moutsopoulos, G.; Simon, J. \emph{Classical Quant. Grav.} \textbf{2009}, \emph{26}, 035016.

\bibitem{Medeiros Hollands} de Medeiros, P.; Hollands, S. \emph{Classical Quant. Grav.} \textbf{2013}, \emph{30}, 175016.

\bibitem{Acik Ertem} A\c{c}{\i}k, \"{O}; Ertem, \"{U}. \emph{Classical Quant. Grav.} \textbf{2015}, \emph{32}, 175007.

\bibitem{Kastor Ray Traschen} Kastor, D.; Ray; S.; Traschen, J. \emph{Classical Quant. Grav.} \textbf{2007}, \emph{24}, 3759-3768.

\bibitem{Benn Charlton} Benn, I. M.; Charlton, P. \emph{Classical Quant. Grav.} \textbf{1997}, \emph{14}, 1037-1042.

\bibitem{Benn Kress} Benn, I. M.; Kress, J. \emph{Classical Quant. Grav.} \textbf{2004}, \emph{21}, 427-431.

\bibitem{Acik Ertem Onder Vercin} A\c{c}{\i}k, \"{O}; Ertem, \"{U}.; \"{O}nder, M.; Ver\c{c}in, A. \emph{Classical Quant. Grav.} \textbf{2009}, \emph{26}, 075001.

\bibitem{Cariglia Krtous Kubiznak} Cariglia, M.; Krtous, P.; Kubiznak, D. \emph{Phys. Rev. D} \textbf{2011}, \emph{84}, 024004.

\bibitem{Papadopoulos} Papadopoulos, G. \emph{Classical Quant. Grav.} \textbf{2008}, \emph{25}, 105016.

\end{references}
\end{document}